\documentclass[pdflatex,sn-mathphys-num, iicol]{sn-jnl}

\usepackage{graphicx}%
\usepackage{multirow}%
\usepackage{amsmath,amssymb,amsfonts}%
\usepackage{amsthm}%
\usepackage{mathrsfs}%
\usepackage[title]{appendix}%
\usepackage{xcolor}%
\usepackage{textcomp}%
\usepackage{manyfoot}%
\usepackage{booktabs}%
\usepackage{algorithm}%
\usepackage{algorithmicx}%
\usepackage{algpseudocode}%
\usepackage{listings}%
\usepackage{siunitx}
\usepackage{soul}
\usepackage{hyperref}

\theoremstyle{thmstyleone}%

\theoremstyle{thmstyletwo}%

\theoremstyle{thmstylethree}%

\begin{document}

\title[title]{All-optical Edge Computing for Speckle Sensing Interrogation}

\author*[1,2]{Tomás Lopes}\email{tomas.j.lopes@inesctec.pt}
\author[1,2]{Joana Teixeira}
\author[1]{Tiago D. Ferreira}
\author[1]{Catarina S. Monteiro}
\author[1,2]{Pedro A. S. Jorge}
\author[1,2]{Nuno A. Silva}


\affil[1]{Centre for Applied Photonics, INESC TEC, Rua do Campo Alegre 687, Porto, 4169-007, Porto, Portugal}

\affil[2]{Department of Physics and Astronomy, Faculty of Sciences, University of Porto, Rua do Campo Alegre s/n, Porto, 4169-007, Porto, Portugal}

\abstract{
Speckle-based sensing exploits the rich environmental information of its high-dimensional spatial intensity patterns. However, the requirement for camera-based acquisition and subsequent electronic digitization introduces significant latency and bandwidth bottlenecks that forbid real-time operation and higher temporal resolutions. Aiming to bypass this imaging processing pipeline, this manuscript presents an optically reconfigurable edge-computing platform for speckle-based sensors that performs task-specific computation directly in the optical domain. This is achieved by projecting output speckle patterns onto a digital micromirror device, using it as a programmable optical layer whose parameters are trained \textit{in situ} using an evolutionary optimization strategy solely from detector feedback. We demonstrate the concept with a multi-point optical fiber sensing task, where multiple piezoelectric actuators simultaneously perturb the fiber, modifying the speckle pattern. Optimizing a set of masks to decouple these concurrent signals, the system successfully achieves real-time signal separation, achieving a target signal enhancement exceeding 4\,dB while suppressing crosstalk leakage below -10\,dB. Operating with bandwidths limited only by the photodetector, this approach paves the way for real-time and ultrafast optical sensing via an all-optical edge computing solution.
}




\maketitle

\footnotetext[1]{Tomás Lopes and Nuno A. Silva contributed equally to this work.}

\section*{Introduction}
\label{sec:intro}

Optical speckle has emerged as an interesting tool for optical sensing and computing, owing to its ability to encode information across a vast number of spatial and modal degrees of freedom through complex interference patterns \cite{Makris2017}. Indeed, the high-dimensional structure of the generated spatial patterns and their sensitivity to external stimuli have been successfully exploited in multimode-fiber sensing \cite{Murray2019, Bennett2020}, optical information processing \cite{Zhu2024}, and computational imaging \cite{Wang2020}. However, the performance of the approach remains fundamentally bounded by the need to record and process full spatial intensity patterns. The bandwidth of the camera sensor and the frame-rate ceilings restrict the temporal resolution, digitization and memory transfer introduce latency, and subsequent processing steps impose computational burdens that scale poorly with dimensionality \cite{Khne2023}. In this context, several alternatives have emerged in recent years, such as the use of event-based vision for speckle readout, which has demonstrated clear performance gains \cite{Gallego2022, lopes2025event}. This trend underscores the growing demand for interrogation architectures capable of operating at high bandwidths without relying on conventional imaging pipelines, and more broadly, extends beyond interrogation alone, reflecting the need for a paradigm shift in sensing and perception systems, as increasing data rates continue to push cloud-centric processing to its limits \cite{Luo2019, Fu2024, skalli2025annealing}. 

To address these limitations, edge computing has emerged as a promising approach, enabling computation to occur at or near the sensor, thereby reducing latency, bandwidth demand, and energy consumption \cite{Caiazza2022}. Importantly, performing computation at the edge also enables alternative sensing modalities - such as event-based vision or high-speed photodetection - that avoid full-frame camera acquisition, thereby significantly increasing the attainable temporal bandwidth compared to conventional imaging hardware \cite{Gallego2022, Delaney2025}. Consistent with this perspective, optical edge computing has been actively investigated for applications including real-time image preprocessing and classification, convolutional operations for computer vision, ultrafast signal filtering and feature extraction in optical communications, and low-latency decision making in sensing systems, where linear and nonlinear transformations are executed directly in the optical domain prior to digitization \cite{Yildirim2024, Hu2024}. However, in the specific context of speckle interrogation, the mappings and operations required for task-specific computation on the output intensity pattern are often complex to model and highly sensitive to physical imperfections. To overcome these challenges, self-adaptive, model-free optimization strategies such as evolutionary algorithms, reinforcement learning, or other gradient-free approaches can be employed to directly program reconfigurable optical elements such as spatial light modulators (SLMs) and metasurfaces, thereby circumventing explicit modeling requirements \cite{Rahmani2018,Lan2025}. 


Leveraging these concepts, this article presents an optically reconfigurable edge-computing solution for the interrogation of speckle-based sensors. The proposed all-optical platform operates directly on speckle patterns by projecting them onto a digital micromirror device (DMD), where task-specific spatial masks are applied to weight different regions of the field prior to detection. To bypass the frame-rate limitations associated with full-field image acquisition, the weighted optical field is instead routed to a pair of differential photodetectors, where square-law detection combined with optical interference produces a nonlinear projection of the encoded information. The masks are optimized using an in-situ training procedure based on a genetic algorithm, where the DMD pattern is optimized using only detector feedback, allowing the system to learn task-specific projections without requiring an explicit physical model of speckle propagation. The capabilities of the proposed scheme are demonstrated for the interrogation of a multi-point acoustic fiber sensing system, showing that perturbations at distinct positions along the fiber can be independently resolved at frequencies in the order of tens of kilohertz. Specifically, our optimized spatial masks effectively decouple the concurrent signals, demonstrating a target signal enhancement exceeding +4 dB and a crosstalk suppression below -10 dB.
Putting in a broader perspective, the results enclosed establish a general framework for optical edge computing in speckle-based sensing, showing how high-dimensional wave information can be compressed and processed at the physical layer. This approach opens new opportunities for high-speed, adaptive sensing across fiber optics, imaging, and wave-based perception systems, where conventional cloud-centric processing is no longer viable.


\section*{Results}
\label{sec:results}

\subsection*{Reconfigurable edge-computing architecture}

To fully take advantage on the high-dimensionality of the output space, conventional interrogation methods of speckle-based sensors need to capture and store the full spatial intensity distribution. In practice, this means that the optical field is an intermediate state that must be digitized and further processed in silico. Although this allows for capitalizing on the benefits of data redundancy and noise averaging, it also translates into a data load that scales with the number of pixels rather than the complexity of the task, imposing bandwidth and latency constraints that restrict real-time operation. Yet, whereas the information of interest is embedded across a high-dimensional interference pattern, the relevant sensing data is often low-dimensional, usually corresponding to the number of independent applied stimuli. This mismatch suggests that, rather than reconstructing the full interference pattern, one can train optical projections to preserve only the information relevant to the sensing task on the detected signal. 

\begin{figure*}
    \centering
    \includegraphics[width=\linewidth]{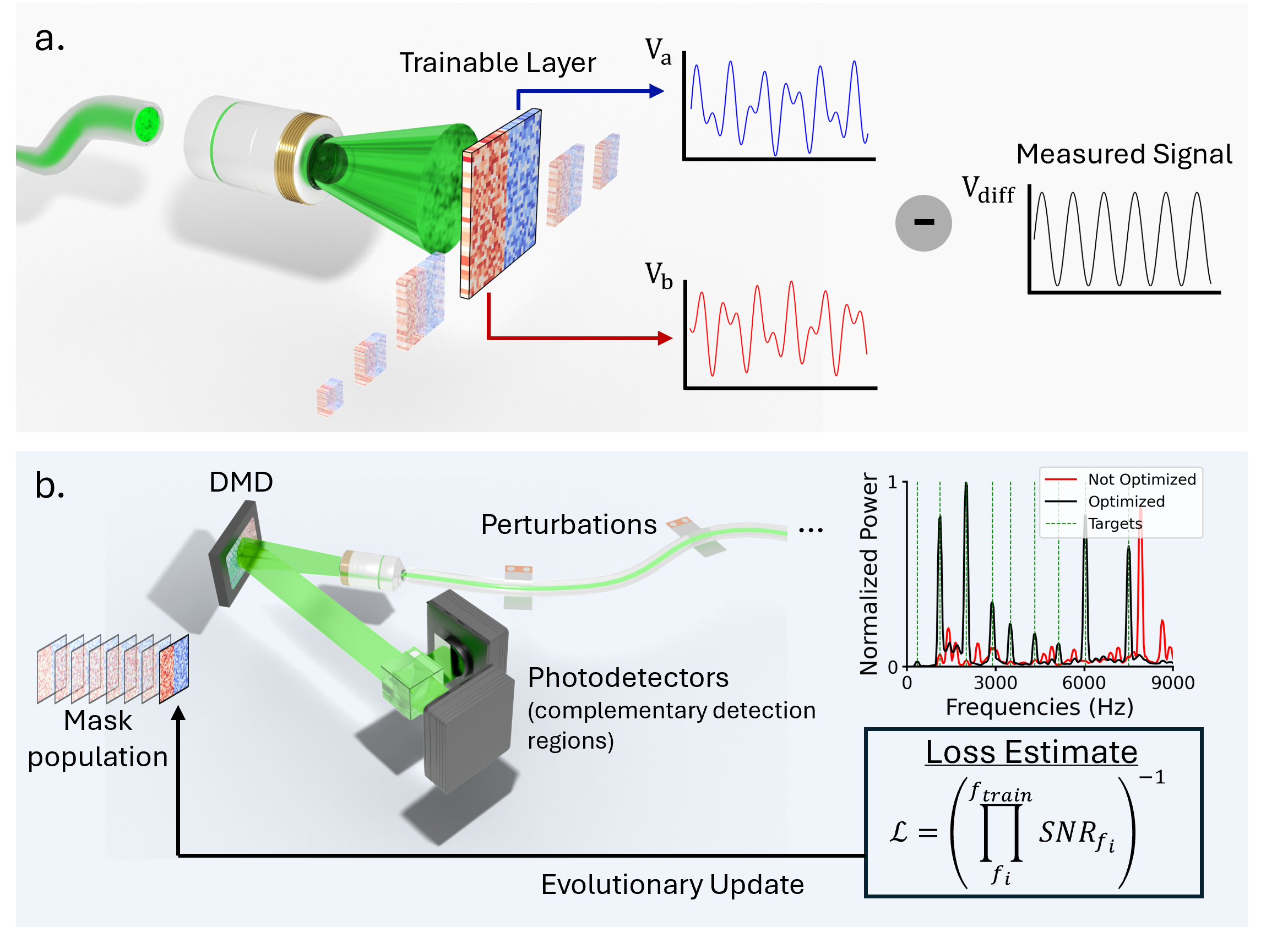}
    \caption{a. Conceptual representation of the readout stage, where a trainable layer modulates the optical field before differential detection. b. Experimental setup illustrating multi-point fiber perturbation via piezoelectric membranes. The output speckle is projected onto a Digital Micromirror Device (DMD) and routed to two photodetectors that capture complementary spatial regions. The training loop demonstrates the iterative optimization process, where a mask population is evaluated against a loss estimate to guide the generation of subsequent populations.}
    \label{fig:main_figure}
\end{figure*}

On this basis, we propose the architecture shown in Figure \ref{fig:main_figure}a, which shifts the mode projection to the optical domain to directly produce readout signals aligned with the desired sensing task.
Conceptually, instead of storing and processing high-dimensional images, the interrogation system uses the speckle field as the input of a programmable optical layer that applies a reconfigurable spatial weight matrix $\boldsymbol{w}$ to map the physical stimuli to the detected output signal. In particular, for the experimental implementation presented, the output speckle was magnified and projected onto a digital micromirror device (DMD), which acted as the programmable layer (see Figure \ref{fig:main_figure}) to modulate the incident field by selectively routing specific pixel regions toward a detection stage. To increase versatility due to the fact that the applied weights on the DMD are necessarily non-negative, we further divided the beam with a beam splitter, and block complementary halves to create two spatially independent parts of the weighted field that are subsequently detected with two distinct photodetectors operating in a differential measurement scheme, $\Delta I =  I_+-I_-$, and positioned in the image plane of the DMD.

In principle, if it is possible to compute a deterministic linear mapping between the applied stimuli and the resulting speckle pattern, one may obtain the optimal map $\boldsymbol{w}$ analytically. Yet, in most cases, the situation is highly complex and non-ideal effects can introduce further challenges (e.g. noise, nonlinearity). In this case, one can still look for the optimal weight map by exploring data-driven methodologies. In particular, in this work, we focus on model-free methods and introduce an in-situ training procedure to program the optical weights. As illustrated in Figure \ref{fig:main_figure}b, we do so by applying a genetic algorithm that iteratively updates the DMD patterns based solely on the feedback from the differential detector voltage, seeking to maximize a task-specific loss function that incentivizes the system to isolate specific sensing events.


\subsection*{Multi-point Sensing and Signal Reconstruction}

In order to demonstrate the proposed concept, we focused on the interrogation and separation of $N$ independent vibrations $\boldsymbol{\delta}=
    \begin{bmatrix}
    \delta_1 & \cdots & \delta_N
    \end{bmatrix}^{\top}$ applied along multiple points of the same optical fiber. 
Assuming that the masks are applied in spatially independent regions and detected only by the corresponding detector as described before, one may model the differential signal in terms of the speckle signal $\boldsymbol{I}(\boldsymbol{\delta})$ at the DMD plane
\begin{equation}
    \Delta I = \boldsymbol{w}_{+}\boldsymbol{I}(\boldsymbol{\delta})^{\top} -\boldsymbol{w}_{-}\boldsymbol{I}(\boldsymbol{\delta})^{\top}.
\end{equation}
In the linear regime, it is possible to show that there exists a set of optimal masks $\boldsymbol{w_{+,-}}$ capable of isolating a given signal, e.g. $\delta_J$, and that it can be computed analytically (see Mathematical Considerations section and supplementary material). Yet, due to the limitations introduced by the DMD and non-idealities of the experimental setup, our work explores instead a data-driven methodology to compute these masks, by maximizing the signal related to $\delta_J$ while minimizing the cross-talk of the other perturbations.

\begin{figure}
    \centering
    \includegraphics[width=\linewidth]{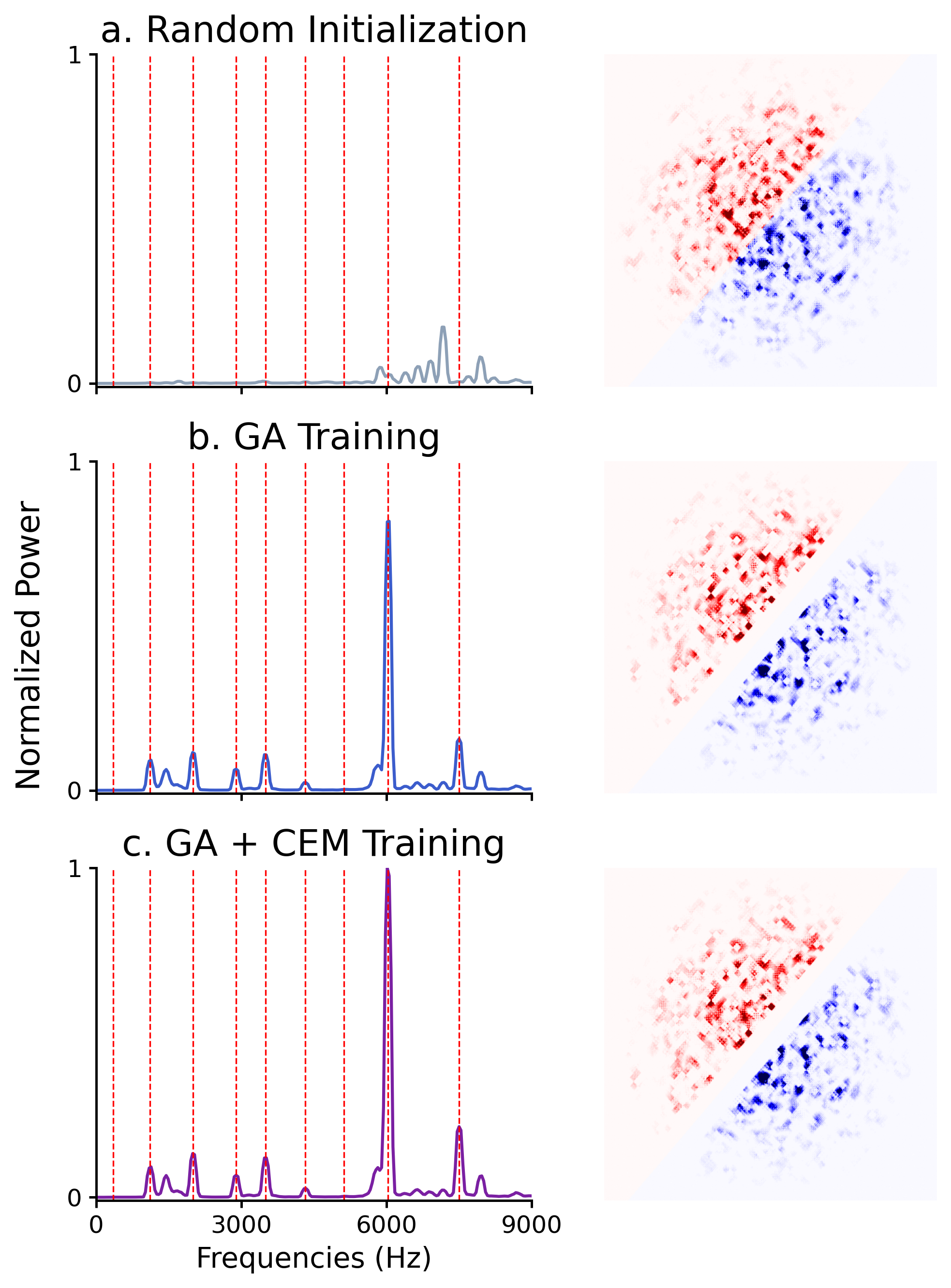}
    \caption{Evolution of the spatial mask optimization process, where the red lines represent the train frequencies used in the target location. a. Random mask initialization stage, where the measured spectrum exhibits weak features and a lack of separation between the target deformation signal and other fiber locations. b. Results following the genetic algorithm training phase, where distinct spectral components emerge at the target frequencies, indicating the successful identification of coarse speckle regions correlated with the intended perturbation. c. Final result of the complete training (genetic algorithm + cross entropy method), which reinforces spatial weights to enhance the signal-to-noise ratio. The right-hand side displays the corresponding speckle patterns modulated by the DMD at each respective stage of the optimization. All these results correspond to the spatial mask optimization process for one perturbed point in the fiber.}
    \label{fig:training_stages}
\end{figure}

To operationalize this methodology, we set up the experiment with three piezoelectric membrane actuators driven by controlled voltage signals at different positions along the fiber (see illustration of Figure \ref{fig:main_figure}). The starting point is the calibration procedure, aiming at identifying the best optical masks to separate the perturbations applied at each point location. For this, we utilize a two-step in-situ optimization training method. First, a genetic algorithm is employed to find a mask that maximizes the power of the signal at target frequencies. This is followed by a fine-tuning of the mask parameters via a cross-entropy method, thereby helping the system converge to an optimal state. As shown in Figure \ref{fig:training_stages}, a predefined set of frequencies is selected as the optimization target and used to define a training loss function based on the signal-ratio, computed directly on the measured signal as
\begin{equation}
    \mathcal{L} = \left( 
    \prod^{\{f_\text{train}\}}_{f_i} \text{SNR}_{f_i} 
    \right)^{-1},
\end{equation}
with
\begin{equation}
    \text{SNR}_{f_i} = \frac{P_{xx}{(f = {f_i})}}{P_{xx}(f \neq {f_\text{train}})}
\end{equation}
where $\{f_\text{train}\}$ is a set of target frequencies exclusive to the location we are optimizing to, and $P_{xx}$ corresponds to the power spectral density of the readout signal. The choice of the inverse function allows for converting the optimization into a minimization problem and is preferred because it provides greater contrast between individuals with high and low SNR, resulting in higher selectiveness in the crossover stage of the genetic algorithm.

\begin{figure}
    \centering
    \includegraphics[width=0.9\linewidth]{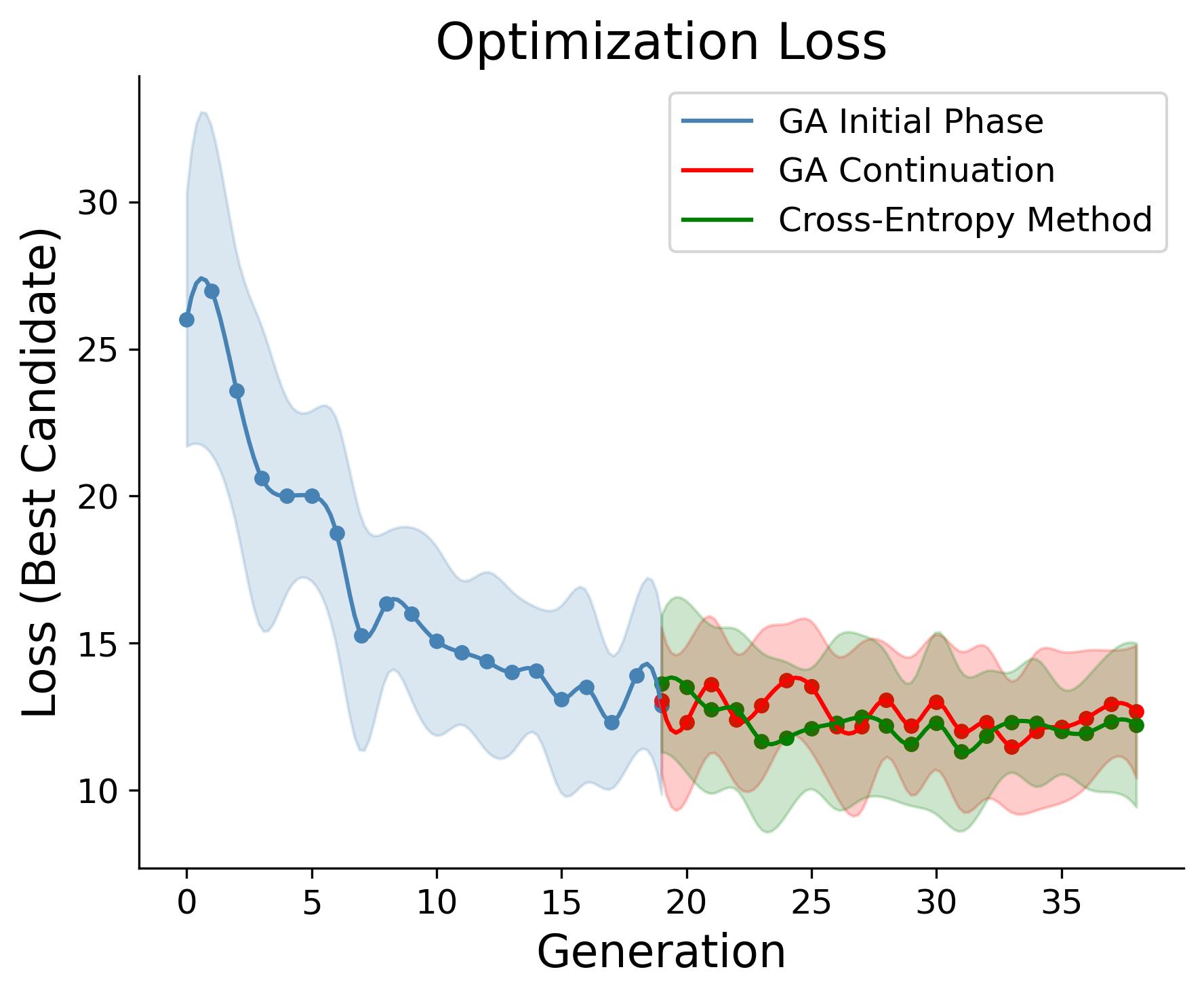}
    \caption{Evolution of loss of the best candidate over the training procedure of one mask. The fluctuations in loss occur because of the frequency shift applied in each generation. The cross-entropy method serves to perform fine-tuning after the initial GA coarse search. The graph is constructed by averaging 10 optimizations.}
    \label{fig:loss_evolution}
\end{figure}

Typical results of the calibration procedure are depicted in Figures \ref{fig:training_stages} and \ref{fig:loss_evolution}. Starting from the results of Figure \ref{fig:training_stages}, it can be seen that the random mask leads to a measured signal whose spectrum exhibits weak and broadly distributed features, with no contrast between target and non-target frequencies, reflecting the absence of task-aligned spatial weighting. After the first optimization with the genetic algorithm, distinct spectral components begin to emerge at the target frequencies, indicating that the algorithm successfully identifies coarse regions of the speckle field that correlate with the intended perturbations. Thus, the genetic algorithm primarily serves to rapidly explore the high-dimensional mask space and establish an informed guess of an interrogation mask for the DMD, albeit with residual background contributions still present. The subsequent cross-entropy stage further improves the separation of the target signal and additional noise by reinforcing the spatial weights associated with the target frequencies, leading to an enhancement of the desired spectral components and increased suppression of off-target energy as seen in the bottom panel of Figure \ref{fig:training_stages}.

\begin{figure*}
    \centering
    \includegraphics[width=\linewidth]{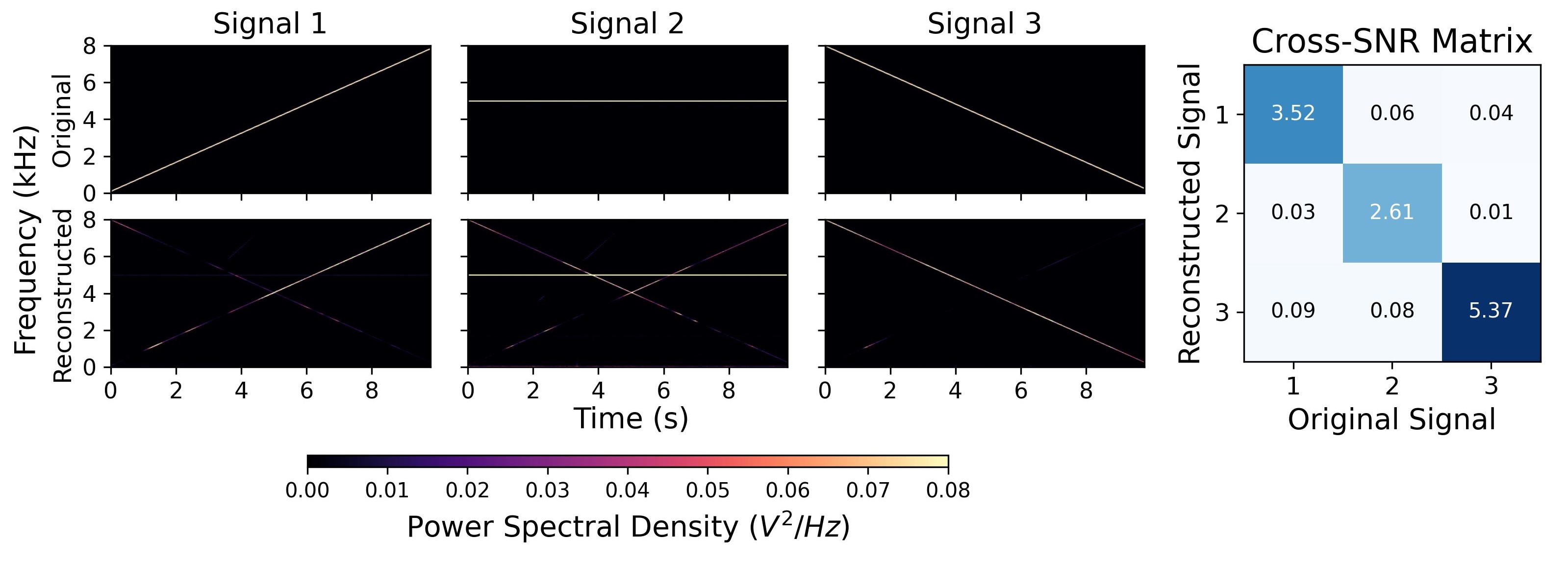}
    \caption{Original and reconstructed spectrograms for three distinct signals playing simultaneously at different deformation points along the optical fiber. Each reconstructed spectrogram is obtained by applying the specific spatial mask optimized to attend to its respective perturbation region. The Cross-SNR matrix quantifies the performance of this approach, confirming that the system successfully isolates the target signal from a mixture of concurrent perturbations with reduced crosstalk.}
    \label{fig:reconstruction_results}
\end{figure*}

After repeating the procedure for each sensing point, the interrogation system is ready to directly retrieve vibration signals applied to the distinct calibrated spatial positions. To detect perturbations at a specific position, one applies the mask optimized for that location and performs an analog measurement of the voltage across the two photodetectors. Typical results are presented in Figure \ref{fig:reconstruction_results} in the form of PSD spectrograms. The top row shows the three concurrent signals, each applied to a different point in the optical fiber, while the bottom row shows the measured signals when the DMD displays the mask targeting that specific perturbation location. From the visual analysis, it is evident that each mask has successfully reduced signal leakage, with a strong representation of the signal playing at the target location. 

To quantify the quality of these results, we introduce a spectrogram-based SNR metric. For each reference signal, we constructed a binary mask for the spectrogram that identifies the time–frequency regions associated with that perturbation. We then computed a Cross-SNR matrix, in which the diagonal elements quantify the fraction of power retained within the correct region, while off-diagonal elements quantify residual leakage into other channels (i.e., crosstalk). Because at some points in time the signal frequencies cross, which can in turn favor the metric by treating leakage as signal, these regions are excluded from the metric, meaning the binary masks only represent exclusive spectral regions.

Let $ P_{R_j} $ denote the power spectrogram of the reconstructed signal $ r_j(t) $, and $ M_i \in \{0,1\} $ the binary mask corresponding to the exclusive regions of reference signal $ s_i(t) $. The Cross-SNR matrix $ A \in \mathbb{R}^{N \times N} $ is defined as:
\begin{equation}
    \text{Cross-SNR}_{i,j} =\frac{\displaystyle \sum M_i  P_{R_j}}{\displaystyle \sum (1 - M_i)P_{R_j}},\qquad i,j = 1, \dots, N \, ,
    \label{eq:cross_snr}
\end{equation}
where the sums are taken over all time-frequency bins of the spectrogram. Note that the denominator includes all spectral energy outside the target region, encompassing both background noise and interference from other sources.

The resulting matrix, shown in Figure \ref{fig:reconstruction_results}(results), exhibits a strong diagonal dominance, providing quantitative validation of the spatial separation capabilities of the system. The stronger diagonal elements prove that the reconstructed signal energy within the target time-frequency footprint is at least two times greater (exceeding $+4$\,dB) than the total residual energy, comprising both background noise and interference from other sources. Conversely, the off-diagonal elements remain consistently below 0.10, proving that the leakage of non-target perturbations into the isolated channels is suppressed to less than one-tenth ($< -10$\,dB) of the residual energy. Together, these values confirm that the optimized spatial masks successfully decouple the complex modal interference, effectively projecting the distinct physical deformations into independent measurements.

\section*{Discussion}
\label{sec:discussion}

This manuscript presents an all-optical reconfigurable edge-computing platform as an interrogation solution for a speckle-based sensor. By projecting the speckle field onto a programmable spatial mask and performing direct differential detection with two photodetectors, the system produces low-dimensional, task-aligned electrical signals at bandwidths limited only by the photodetectors, rather than by sensor frame rates or electronic processing pipelines. In specific, we introduce an in-situ evolutionary optimization framework to learn the optical masks directly from performance feedback, enabling the system to adapt to the complex and highly nonlinear mapping between physical perturbations and speckle patterns. 

To demonstrate the concept and its practical utility, we used it to separate multiple perturbations applied along a multimode optical fiber. In the context of the present work, this problem is particularly relevant to illustrate the limitations of standard camera-based interrogation approaches. Indeed, although such systems have previously demonstrated the capacity to detect and interrogate these perturbations \cite{Cai2023,Gu2023}, their performance is fundamentally limited by the frame rate of the sensor and the need for digital post-processing. This manifests itself not only in memory bandwidth limitations but also in significant latency, restricting the ability of the system to track fast, dynamic, and high-frequency changes (at most a few hundred Hz). In contrast, our optical edge-computing framework allows us to separate these perturbations directly in the analog domain, effectively bypassing the camera bottleneck entirely. This is enabled by the use of single-pixel detectors, which can operate at bandwidths up to MHz and even GHz, thereby supporting real-time, high-speed interrogation. 


The results presented, specifically the spectrogram-based Cross-SNR analysis, confirm that the optimized masks successfully isolate the target perturbations while suppressing residual contributions from other sources, and the real-time audio capability further highlights the practical implications of this approach. Because the separation occurs optically, the output signal is immediately available as a high-bandwidth analog voltage, allowing for direct interfacing with downstream hardware without the need for intermediate digitization or computation.  This means that the analog output voltage signal is ready for immediate readout and action. Additional results regarding the measured readout signals can be found in the supplemental material, where signals of higher spectral complexity were tested.

In this manuscript, these capabilities were demonstrated in the audible range, up to around 8\,kHz; however, it is important to note that this bandwidth ceiling occurs mostly from the mechanical coupling of the vibrations to the optical fiber (which can be improved using tailored structural designs such as mandrel-like structures\cite{uyar2025compliant, lima2010mandrel}), rather than a limitation at the end of the training framework. Besides, while the present work focuses on the separation of distributed perturbations along a fiber, the same principles naturally extend to other speckle-based tasks where information is encoded in high-dimensional interference patterns. For instance, the platform could be adapted to discriminate particles or defects using speckle signatures\cite{wang2023specklenn, castilho2023machine, newaz2023machine}, with the optical masks being trained to respond selectively to specific classes rather than spatial locations, enabling real-time and high-throughput classification capabilities and establishing a promising route for novel applications.

\section*{Methods}
\label{sec:methods}

\subsection*{Evolutionary Algorithms}

Our architecture relies on evolutionary algorithms to optimize the DMD masks directly from the detector feedback, and so, we apply a two-step strategy involving a genetic algorithm for global exploration, followed by the cross-entropy method for fine-tuning.

The genetic algorithm begins with a set of random initial masks. These are updated over a predefined number of generations, where at each iteration, we apply a loss function that identifies the ones with the highest SNR for the target perturbation while suppressing others. A portion of the population with the lowest loss scores is selected, after which we apply tournament selection, crossover, and mutation. This allows the system to converge to a robust initial guess of the optimal interrogation mask.

To perform a finer optimization, we continue with the Cross-Entropy method, using the final output of the genetic algorithm to initialize the starting population. In this stage, each macropixel is treated as an independent probability distribution. We generate a new population by sampling from these distributions and evaluating their loss using the same metric as before. Based on the best-performing samples (elites), we update the probability parameters to steer the distributions toward the optimal solution. For both training phases, we use a batch size of 35 masks per training iteration, with GA training for 20 generations and CEM for 15.

\subsection*{Mathematical Considerations}

To understand the interrogation concept related to the separation of the perturbations, we start by analyzing the signal $I_{out}$ obtained for a bucket detector positioned at the image plane of the DMD. In the perturbative linear regime (see the supplementary material for further discussion of the linearity), we can approximate it by 
\begin{equation}
I_{\text{out}} \approx \boldsymbol{w}I|_{\boldsymbol{\delta}=0}^{\top} 
+ \sum_{i=1}^{N}  \boldsymbol{w}\left( \delta_i\partial_{\delta_i}\boldsymbol{I}\right)|_{\boldsymbol{\delta}=0}^{\top},
\end{equation}
where $\boldsymbol{I}:=\boldsymbol{I}(\boldsymbol{\delta})$ is a function of the perturbations and corresponds to the intensity pattern at the DMD plane, and $\boldsymbol{w}$ is a mask applied to it. Aiming to isolate the perturbation at a given index $J$, we may construct a system of $N$ equations
\begin{equation}
    \boldsymbol{w}^{(J)}\left( \partial_{\delta_i}\boldsymbol{I}\right)|_{\boldsymbol{\delta}=0}^{\top}= \delta_{iJ} \label{eq:system}
\end{equation}
being $\delta_{iJ}$ the discrete Kronecker delta. Algebraically, it can be shown that the system has solution $\boldsymbol{w}^{(J)}$ if the sensitivities $\left( \partial_{\delta_i}\boldsymbol{I}\right)|_{\boldsymbol{\delta}=0}$ (whose dimension corresponds to the number of pixels utilized on the DMD) are linearly independent, which is usually true as long as the number of spatial modes of the fiber is much larger than $N$. 

While this proves that it is possible to interrogate each of the perturbations individually by computing an optimal interrogation mask $\boldsymbol{w}^{(J)}$, we should also note that the use of a DMD introduces additional constraints to the solvability of the equation system \ref{eq:system} as the weights of $\boldsymbol{w}^{(J)}$ are necessarily non-negative. In order to bypass this limitation, we utilize two photodetectors working in differential mode, $I_{out} = I_+-I_-$, each one imaging two spatially independent zones associated with trainable masks $\boldsymbol{w_{+,-}}^{(J)}$, respectively. Although not guaranteed upfront, one may still argue that if the dimension of the pixel space corresponding to each mask is sufficiently large compared to the number of perturbations (specifically, larger than $N^2$ \cite{dremeau2015reference, Popoff2010}), then probabilistically, all the information will be present in each of the masks due to redundancy, which goes back to the scenario described above.

\subsection*{Optical Setup}

\begin{figure}[H]
    \centering
    \includegraphics[width=\linewidth]{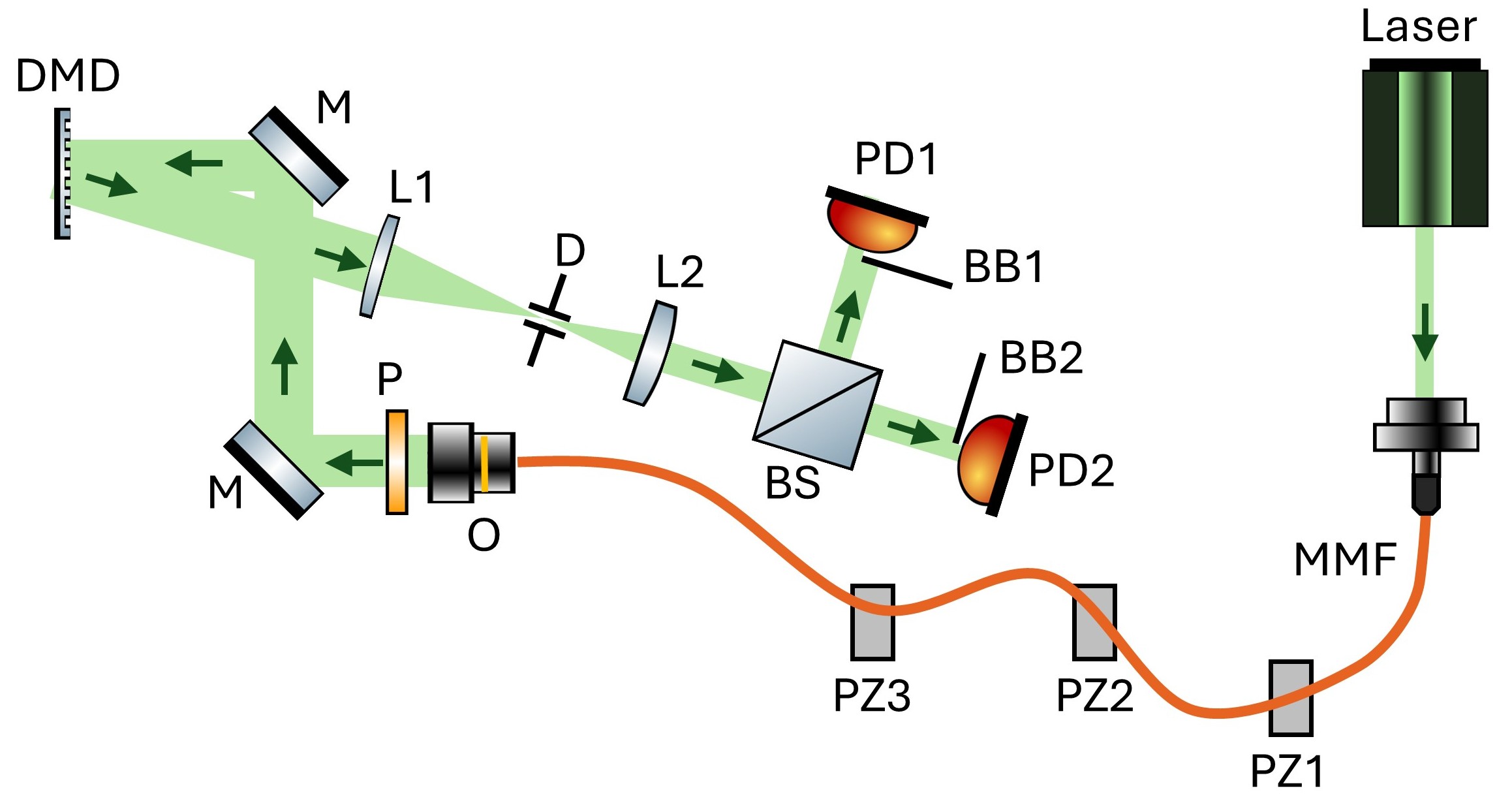}
    \caption{DMD, digital micromirror device; L$_i$, lenses; P, polarizer; M, mirrors, PD$_i$, photodetectors; BB$_i$, beam blockers; O, objective; D, diaphragm; PZ$_i$, piezelectric actuators; MMF, multimode optical fiber; BS, beam splitter.}
    \label{fig:schematic_setup}
\end{figure}

The core optical setup is driven by a 532\,nm continuous-wave laser (CNI, MSL-DS-532), coupled into a 6\,m segment of multimode, 50\, $\mu$m core, optical fiber. The output speckle field is subsequently magnified and collimated by a 10$\times$ microscope objective, followed by a polarizer, and then projected onto a digital micromirror device (Vialux, SuperSpeed V-Module V-7000 VIS). Because both evolutionary algorithms optimize over a continuous range of weight values, and the DMD is inherently a binary display device, grayscale levels are implemented spatially by grouping DMD pixels into $4 \times 4$ macropixels, around the same size as the speckle grain. Within each block, a Bayer dithering scheme is applied, yielding 16 discrete intensity levels. The encoded light field is imaged using a 4F imaging system using 200\,mm and a 100\,mm plano convex lens providing a 2$\times$ reduction. The light is then split 50/50 using a beam splitter, and in each path, complementary regions of the light field are blocked just before entering two photodetectors (Thorlabs, PDA100A and PDA36A), which are connected to a data acquisition system (National Instruments, 6259-USB) that performs differential measurements across two analog terminals. In addition to performing analog measurements, the data acquisition system is also responsible for sending analog signals to three piezoelectric membranes (TDK, PHUA3015-049B-00-000). For real-time signal playback, the differential measurement is connected to a TDA 2050 audio amplifier, which is then connected to a speaker to produce directly audible signals.

\section*{Acknowledgments}

Tomás Lopes and Joana Teixeira acknowledge the support of the Foundation for Science and Technology (FCT), Portugal, through Grants 2024.01830.BD and 2024.00426.BD, respectively. Nuno A. Silva acknowledges the support of FCT under the
grant 2022.08078.CEECIND/CP1740/CT0002. Tiago D. Ferreira acknowledges the support of FCT under the grant 2024.10684.CEECIND

\bibliography{sn-bibliography}

@article{Makris2017,
   author = {Konstantinos G. Makris and Andre Brandstötter and Philipp Ambichl and Ziad H. Musslimani and Stefan Rotter},
   doi = {10.1038/lsa.2017.35},
   issn = {20477538},
   issue = {9},
   journal = {Light: Science and Applications},
   month = {9},
   publisher = {Nature Publishing Group},
   title = {Wave propagation through disordered media without backscattering and intensity variations},
   volume = {6},
   year = {2017}
}

@article{Popoff2010,
   author = {S. M. Popoff and G. Lerosey and R. Carminati and M. Fink and A. C. Boccara and S. Gigan},
   doi = {10.1103/PhysRevLett.104.100601},
   issn = {00319007},
   issue = {10},
   journal = {Physical Review Letters},
   month = {3},
   pmid = {20366410},
   title = {Measuring the transmission matrix in optics: An approach to the study and control of light propagation in disordered media},
   volume = {104},
   year = {2010}
}

@article{Bennett2020,
   author = {Aviya Bennett and Yevgeny Beiderman and Sergey Agdarov and Yafim Beiderman and Rotem Hendel and Barak Straussman and Zeev Zalevsky},
   doi = {10.1364/oe.384423},
   issn = {10944087},
   issue = {14},
   journal = {Optics Express},
   month = {7},
   pages = {20830},
   pmid = {32680135},
   publisher = {Optica Publishing Group},
   title = {Monitoring of vital bio-signs by analysis of speckle patterns in a fabric-integrated multimode optical fiber sensor},
   volume = {28},
   year = {2020}
}

@article{Murray2019,
   author = {Matthew J. Murray and Allen Davis and Clay Kirkendall and Brandon Redding},
   doi = {10.1364/oe.27.028494},
   issn = {10944087},
   issue = {20},
   journal = {Optics Express},
   month = {9},
   pages = {28494},
   pmid = {31684600},
   publisher = {The Optical Society},
   title = {Speckle-based strain sensing in multimode fiber},
   volume = {27},
   year = {2019}
}

@article{Zhu2024,
   author = {Yusi Zhu and Zhaoke Mi and Yupeng Zhu and Changjun Ke and Lu Rong and Yishi Shi},
   doi = {10.1016/j.optcom.2024.131022},
   issn = {00304018},
   journal = {Optics Communications},
   month = {12},
   publisher = {Elsevier B.V.},
   title = {Optical information hiding based on speckle encoding with dual-multiplexing interferometry},
   volume = {573},
   year = {2024}
}

@article{Wang2020,
   author = {Xiaoxia Wang and Yong Tao and Fengbao Yang and Yiwei Zhang},
   doi = {10.1016/j.optcom.2019.124470},
   issn = {00304018},
   journal = {Optics Communications},
   month = {1},
   publisher = {Elsevier B.V.},
   title = {An effective compressive computational ghost imaging with hybrid speckle pattern},
   volume = {454},
   year = {2020}
}

@article{Gallego2022,
   author = {Guillermo Gallego and Tobi Delbruck and Garrick Orchard and Chiara Bartolozzi and Brian Taba and Andrea Censi and Stefan Leutenegger and Andrew J. Davison and Jorg Conradt and Kostas Daniilidis and Davide Scaramuzza},
   doi = {10.1109/TPAMI.2020.3008413},
   issn = {19393539},
   issue = {1},
   journal = {IEEE Transactions on Pattern Analysis and Machine Intelligence},
   month = {1},
   pages = {154-180},
   pmid = {32750812},
   publisher = {IEEE Computer Society},
   title = {Event-Based Vision: A Survey},
   volume = {44},
   year = {2022}
}

@inproceedings{Khne2023,
   author = {Jonas Kühne and Michele Magno and Luca Benini},
   doi = {10.1109/IWASI58316.2023.10164626},
   isbn = {979-8-3503-3694-8},
   booktitle = {2023 9th International Workshop on Advances in Sensors and Interfaces (IWASI)},
   month = {6},
   pages = {143-148},
   publisher = {IEEE},
   title = {A Fast and Accurate Optical Flow Camera for Resource-Constrained Edge Applications},
   url = {https://ieeexplore.ieee.org/document/10164626/},
   year = {2023}
}

@article{Luo2019,
   author = {Yi Luo and Deniz Mengu and Nezih T. Yardimci and Yair Rivenson and Muhammed Veli and Mona Jarrahi and Aydogan Ozcan},
   doi = {10.1038/s41377-019-0223-1},
   issn = {20477538},
   issue = {1},
   journal = {Light: Science and Applications},
   month = {12},
   publisher = {Springer Nature},
   title = {Design of task-specific optical systems using broadband diffractive neural networks},
   volume = {8},
   year = {2019}
}

@misc{Fu2024,
   author = {Tingzhao Fu and Jianfa Zhang and Run Sun and Yuyao Huang and Wei Xu and Sigang Yang and Zhihong Zhu and Hongwei Chen},
   doi = {10.1038/s41377-024-01590-3},
   issn = {20477538},
   issue = {1},
   journal = {Light: Science and Applications},
   month = {12},
   publisher = {Springer Nature},
   title = {Optical neural networks: progress and challenges},
   volume = {13},
   year = {2024}
}

@article{Caiazza2022,
   author = {Chiara Caiazza and Silvia Giordano and Valerio Luconi and Alessio Vecchio},
   doi = {10.1016/j.comcom.2022.07.026},
   issn = {1873703X},
   journal = {Computer Communications},
   month = {10},
   pages = {213-225},
   publisher = {Elsevier B.V.},
   title = {Edge computing vs centralized cloud: Impact of communication latency on the energy consumption of LTE terminal nodes},
   volume = {194},
   year = {2022}
}

@article{Delaney2025,
   author = {Ethan Delaney and Tim Brophy and Enda Ward and Fiachra Collins and Edward Jones and Brian Deegan and Martin Glavin},
   doi = {10.1109/JSEN.2025.3584460},
   issn = {15581748},
   issue = {16},
   journal = {IEEE Sensors Journal},
   pages = {31545-31562},
   publisher = {Institute of Electrical and Electronics Engineers Inc.},
   title = {Evaluating Event-Based Vision Sensing in Rain and Fog},
   volume = {25},
   year = {2025}
}

@article{Yildirim2024,
   author = {Mustafa Yildirim and Niyazi Ulas Dinc and Ilker Oguz and Demetri Psaltis and Christophe Moser},
   doi = {10.1038/s41566-024-01494-z},
   issn = {17494893},
   issue = {10},
   journal = {Nature Photonics},
   month = {10},
   pages = {1076-1082},
   publisher = {Nature Research},
   title = {Nonlinear processing with linear optics},
   volume = {18},
   year = {2024}
}

@article{Hu2024,
   author = {Jingtian Hu and Kun Liao and Niyazi Ulas Dinç and Carlo Gigli and Bijie Bai and Tianyi Gan and Xurong Li and Hanlong Chen and Xilin Yang and Yuhang Li and Çağatay Işıl and Md Sadman Sakib Rahman and Jingxi Li and Xiaoyong Hu and Mona Jarrahi and Demetri Psaltis and Aydogan Ozcan},
   doi = {10.1186/s43593-024-00067-5},
   issn = {26628643},
   issue = {1},
   journal = {eLight},
   month = {12},
   publisher = {Springer},
   title = {Subwavelength imaging using a solid-immersion diffractive optical processor},
   volume = {4},
   year = {2024}
}

@article{Rahmani2018,
   author = {Babak Rahmani and Damien Loterie and Georgia Konstantinou and Demetri Psaltis and Christophe Moser},
   doi = {10.1038/s41377-018-0074-1},
   issn = {20477538},
   issue = {1},
   journal = {Light: Science and Applications},
   month = {12},
   publisher = {Nature Publishing Group},
   title = {Multimode optical fiber transmission with a deep learning network},
   volume = {7},
   year = {2018}
}

@article{Lan2025,
   author = {Ruibo Lan and Yang Li and Hanyang Shen and Yuxin Lu and Hongbin Hu and Longjun Zheng and Qiang Zong and Yubin Zang and Zuxing Zhang},
   doi = {10.1364/oe.571745},
   issn = {10944087},
   issue = {17},
   journal = {Optics Express},
   month = {8},
   pages = {36133},
   pmid = {40984386},
   publisher = {Optica Publishing Group},
   title = {All-optical color image encryption using multimode fiber speckles and diffractive deep neural networks},
   volume = {33},
   year = {2025}
}

@article{Cai2023,
   author = {Yuezhi Cai and Yan Liu and Guangde Li and Qi Qin and Lezhi Pang and Wenhua Ren and Jie Wei and Muguang Wang},
   doi = {10.1016/j.optlastec.2022.109062},
   issn = {00303992},
   journal = {Optics and Laser Technology},
   month = {5},
   publisher = {Elsevier Ltd},
   title = {Reflective tactile sensor assisted by multimode fiber-based optical coupler and fiber specklegram},
   volume = {160},
   year = {2023}
}

@article{Gu2023,
   author = {Liangliang Gu and Han Gao and Haifeng Hu},
   doi = {10.3390/nano13040768},
   issn = {20794991},
   issue = {4},
   journal = {Nanomaterials},
   month = {2},
   publisher = {MDPI},
   title = {Demonstration of a Learning-Empowered Fiber Specklegram Sensor Based on Focused Ion Beam Milling for Refractive Index Sensing},
   volume = {13},
   year = {2023}
}

@article{lopes2025event,
  title={Event-based Speckle Interrogation for High-Bandwidth Multi-point Optical Fiber Sensing},
  author={Lopes, Tom{\'a}s and Teixeira, Joana and Rocha, Vicente V and Ferreira, Tiago D and Monteiro, Catarina S and Jorge, Pedro AS and Silva, Nuno A},
  journal={arXiv preprint arXiv:2509.22094},
  year={2025}
}

@article{skalli2025annealing,
  title={Annealing-inspired training of an optical neural network with ternary weights},
  author={Skalli, Anas and Goldmann, Mirko and Haghighi, Nasibeh and Reitzenstein, Stephan and Lott, James A and Brunner, Daniel},
  journal={Communications Physics},
  volume={8},
  number={1},
  pages={68},
  year={2025},
  publisher={Nature Publishing Group UK London}
}

@article{uyar2025compliant,
  title={A Compliant Mandrel-Based Fiber Optic Hydrophone for Underwater Acoustic Sensing},
  author={Uyar, Faruk and Aldemir, A Safa and Kartaloglu, Tolga and Ozbay, Ekmel and Ozdur, Ibrahim},
  journal={IEEE Sensors Journal},
  year={2025},
  publisher={IEEE}
}

@article{lima2010mandrel,
  title={Mandrel-based fiber-optic sensors for acoustic detection of partial discharges—A proof of concept},
  author={Lima, Sanderson EU and Fraz{\~a}o, Orlando and Farias, Rubem G and Ara{\'u}jo, Francisco M and Ferreira, Luis A and Santos, Jos{\'e} L and Miranda, Vladimiro},
  journal={IEEE Transactions on Power Delivery},
  volume={25},
  number={4},
  pages={2526--2534},
  year={2010},
  publisher={IEEE}
}

@article{wang2023specklenn,
  title={SpeckleNN: a unified embedding for real-time speckle pattern classification in X-ray single-particle imaging with limited labeled examples},
  author={Wang, Cong and Florin, Eric and Chang, H-Y and Thayer, Jana and Yoon, Chun Hong},
  journal={IUCrJ},
  volume={10},
  number={5},
  pages={568--578},
  year={2023},
  publisher={International Union of Crystallography}
}

@article{castilho2023machine,
  title={Machine learning classification of speckle patterns for roughness measurements},
  author={Castilho, VM and Balthazar, WF and da Silva, L and Penna, TJP and Huguenin, JAO},
  journal={Physics Letters A},
  volume={468},
  pages={128736},
  year={2023},
  publisher={Elsevier}
}

@article{newaz2023machine,
  title={Machine-learning-enabled multimode fiber specklegram sensors: a review},
  author={Newaz, Asif and Faruque, Md Omar and Al Mahmud, Rabiul and Sagor, Rakibul Hasan and Khan, Mohammed Zahed Mustafa},
  journal={IEEE Sensors Journal},
  volume={23},
  number={18},
  pages={20937--20950},
  year={2023},
  publisher={IEEE}
}

@article{dremeau2015reference,
  title={Reference-less measurement of the transmission matrix of a highly scattering material using a DMD and phase retrieval techniques},
  author={Dr{\'e}meau, Ang{\'e}lique and Liutkus, Antoine and Martina, David and Katz, Ori and Sch{\"u}lke, Christophe and Krzakala, Florent and Gigan, Sylvain and Daudet, Laurent},
  journal={Optics express},
  volume={23},
  number={9},
  pages={11898--11911},
  year={2015},
  publisher={Optical Society of America}
}

\end{document}


\title[title]{All-optical Edge Computing for Speckle Sensing Interrogation - Supplementary Material}

\author*[1,2]{\fnm{Tomás} \sur{Lopes}}\email{tomas.j.lopes@inesctec.pt}
\equalcont{These authors contributed equally to this work.}

\author[1,2]{\fnm{Joana} \sur{Teixeira}}

\author[1]{\fnm{Tiago} \sur{D. Ferreira}}

\author[1]{\fnm{Catarina} \sur{S.  Monteiro}}

\author[1,2]{\fnm{Pedro} \sur{A. S. Jorge}}

\author[1,2]{\fnm{Nuno} \sur{A. Silva}}
\equalcont{These authors contributed equally to this work.}

\affil[1]{\orgdiv{Centre for Applied Photonics}, \orgname{INESC TEC}, \orgaddress{\street{Rua do Campo Alegre 687}, \city{Porto}, \postcode{4169-007}, \state{Porto}, \country{Portugal}}}

\affil[2]{\orgdiv{Department of Physics and Astronomy}, \orgname{Faculty of Sciences, University of Porto}, \orgaddress{\street{Rua do Campo Alegre s/n}, \city{Porto}, \postcode{4169-007}, \state{Porto}, \country{Portugal}}}

\maketitle

\section{Speckle Dynamics in Multimode Optical Fibers}
\label{sec:intro}

Multimode optical fibers (MMF) support a large number of spatial modes, each characterized by unique propagation constants and phase velocities. The coherent superposition of these modes at the fiber output results in a complex interference pattern referred to as speckle. While spatially complex, the speckle field is a deterministic function of the physical state of the fiber. Environmental perturbations—such as mechanical vibrations from piezoelectric actuators—alter the relative modal phases and intermodal coupling, leading to a modulation of the output speckle pattern. Using the transmission matrix formalism, the output electric field $\mathbf{E_{out}}$ in the presence of a localized perturbation $\delta$ at a specific position $x$ along the fiber can be expressed as:

\begin{equation}
    \mathbf{E_{out}}(\delta) = \mathbf{M_2 D}(\delta) \mathbf{M_1} E_{in},
    \label{eq:output_field},
\end{equation}

where $\mathbf{M_1}$ represents the transmission matrix from the source to the perturbation site, $\mathbf{D}(\delta)$ describes the local change in propagation due to the deformation, and $\mathbf{M_2}$ represents the transmission through the remaining fiber length.

\subsection{Linearization of the Perturbation Regime}

In the regime of small perturbations ($\delta \ll 1$), we can perform a first-order Taylor expansion of the deformation matrix around the unperturbed state $\mathbf{D_0}$:

\begin{equation}
    \mathbf{D}(\delta) \approx \mathbf{D_0} + \left( \frac{\partial \mathbf{D}}{\partial \delta} \right) \delta.
    \label{eq:perturbation_expansion}
\end{equation}

Since, for most practical purposes, optical detection relies on square-law detectors, we consider the output intensity $I(\delta) = |\mathbf{E_{out}}|^2$. Substituting Eq. \ref{eq:perturbation_expansion} into Eq. \ref{eq:output_field} and neglecting higher-order terms $\mathcal{O}(\delta^2)$, the intensity at a given spatial coordinate $x$ evolves linearly with the perturbation:

\begin{equation}
    I(x; \delta) \approx I_0(x) + \left( \frac{\partial I(x)}{\partial \delta} \right) \delta,
    \label{eq:output_intensity}
\end{equation}

where $I_0(x)$ is the static speckle intensity. For $N$ independent perturbations $\{\delta_i\}$ occurring at different positions $\{z_i\}$ along the fiber, the total change in the speckle pattern $\Delta \mathbf{I} = \mathbf{I} - \mathbf{I}_0$ can be approximated as a linear superposition of the individual contributions:

\begin{equation}
    \Delta \mathbf{I} \approx \sum_{i=1}^N \mathbf{J}_i \delta_i = \mathbf{J} \boldsymbol{\delta},
\end{equation}

where $\mathbf{J}_i = \partial \mathbf{I} / \partial \delta_i$ is the sensitivity vector (or Jacobian) for the $i$-th perturbation site, and $\mathbf{J}$ is the sensitivity matrix.

\subsection{Separability via Linear Projection}

The existence of a linear relationship between physical perturbations and the resulting speckle intensity variations implies that the information regarding each source $\delta_i$ is linearly embedded within the high-dimensional space of the speckle pattern. This linear structure ensures that there is a set of optimal spatial weights capable of isolating a target signal from a mixture of perturbations. In this context, the speckle field serves as a high-dimensional input to a programmable optical layer that maps physical changes in the fiber onto discrete detector measurements.

Because the relevant sensing data is often low-dimensional compared to the high-dimensional interference pattern, a linear interrogation layer can effectively reduce the dimensionality of the system while performing task-specific computation. By selecting specific spatial weights, the system can selectively filter the optical field to isolate perturbations occurring at specific fiber segments. This theoretical framework demonstrates that complex distributed sensing data can be decoupled and reduced to low-dimensional signals in the analog domain, bypassing the need for full-field imaging or digital post-processing. Although the exact mapping is highly sensitive to physical imperfections, the vast number of spatial and modal degrees of freedom in the fiber ensures that a solution for signal separation remains physically accessible within the speckle field.

\section{Linearization of the transmission matrix}

A key aspect of the model described here is the linearization of the transmission matrix, provided that a sufficiently small deformation is applied (equation \ref{eq:perturbation_expansion}). In this regime, the perturbation acts as a first-order correction to the unperturbed transmission matrix, such that increasing the applied voltage only scales the amplitude of a fixed transmission modification. Consequently, the resulting speckle intensity variations should preserve their spatial structure and evolve proportionally with the applied deformation.

\begin{figure}[H]
    \centering
    \includegraphics[width=\linewidth]{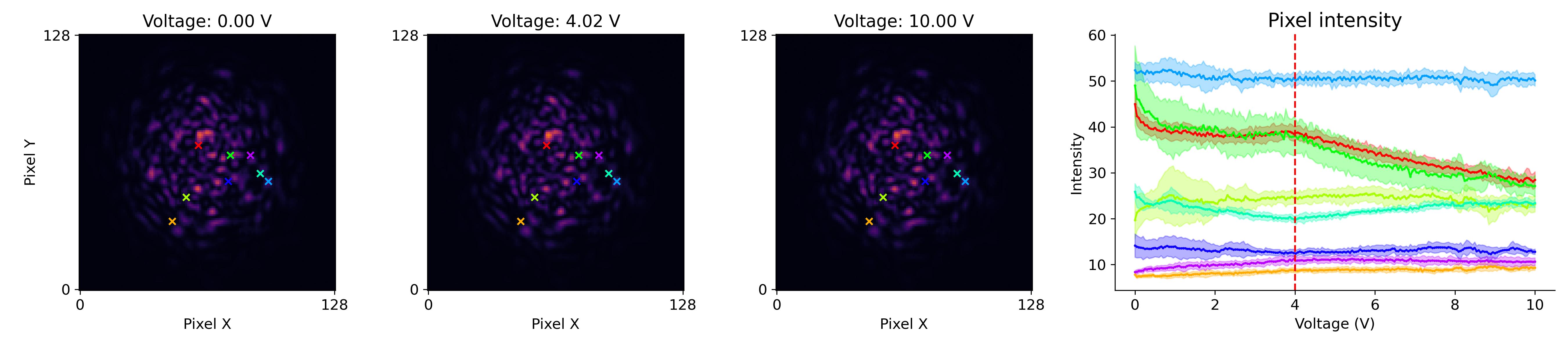}
    \caption{The three images on the left demonstrate the evolution of the speckle pattern with the increasing voltage. The image on the right demonstrates the evolution of the intensity for eight randomly selected pixels.}
    \label{fig:matrix_linearization}
\end{figure}

To experimentally verify the validity of this approximation, a sequence of constant voltages ranging from 0 to 10\,V was applied to a single piezoelectric membrane, and the corresponding output speckle patterns were recorded at each step (Figure \ref{fig:matrix_linearization}).

\begin{figure}[H]
    \centering
    \includegraphics[width=0.35\linewidth]{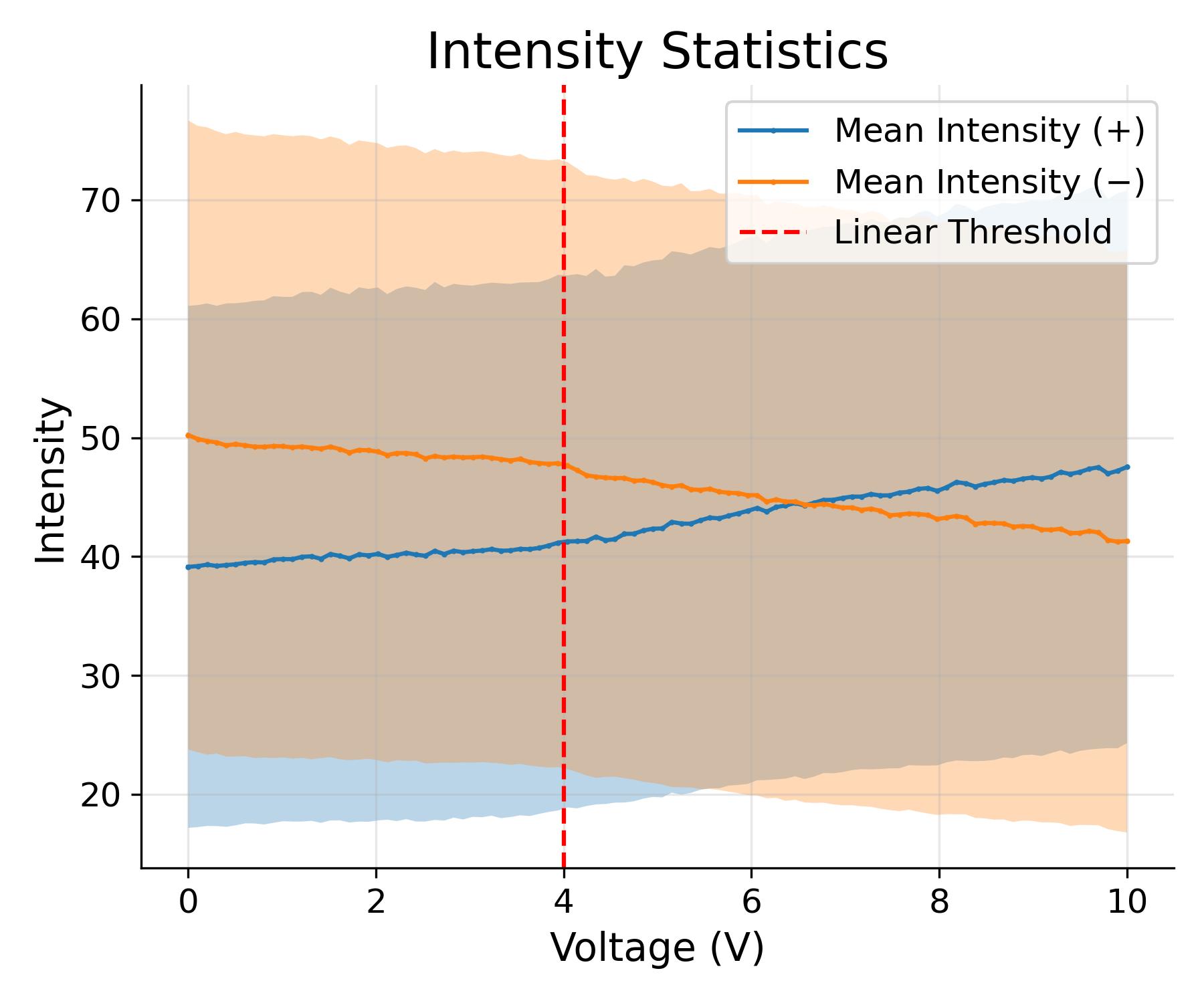}
    \caption{Intensity as a function of the applied voltage. Here we separate pixels that increase and decrease intensity and average them.}
    \label{fig:matrix_linearization_global}
\end{figure}

For applied voltages up to approximately 4\,V, the speckle evolution exhibits a consistent spatial response, with intensity variations across the pattern scaling proportionally with voltage while maintaining the same overall structure. This behavior indicates that the deformation remains within the small-perturbation regime, where the transmission matrix is well described by a first-order expansion.
We can observe this behaviour globally by separating the pixels that increase and decrease their intensity over the applied deformation and plotting them as seen in Figure \ref{fig:matrix_linearization_global}

Above 4\,V, a clear change in the evolution is observed, marked by a slope change in the intensity, which coincides with a modification of the spatial response pattern, indicating that higher-order perturbation terms become significant. In this regime, the deformation no longer acts as a simple scaling of a single transmission perturbation, and the linearized model no longer fully describes the system.

\subsection{Cross Terms}

Another central assumption of the model is that second-order cross terms ($\mathcal{O}(\delta_i\delta_j)$) can be neglected within the operating regime. To assess this experimentally, we optimized a mask for a specific deformation location by driving it at $f_0 = 4000\,Hz$ and holding the voltage fixed ($0.1$, $0.2$, and $0.3\,V$ in three independent runs). The remaining channels were driven at a different frequency, $f_1 = 1341\,Hz$, while their voltage was swept from $0$ to $1\,V$.

\begin{figure}[H]
    \centering
    \includegraphics[width=.9\linewidth]{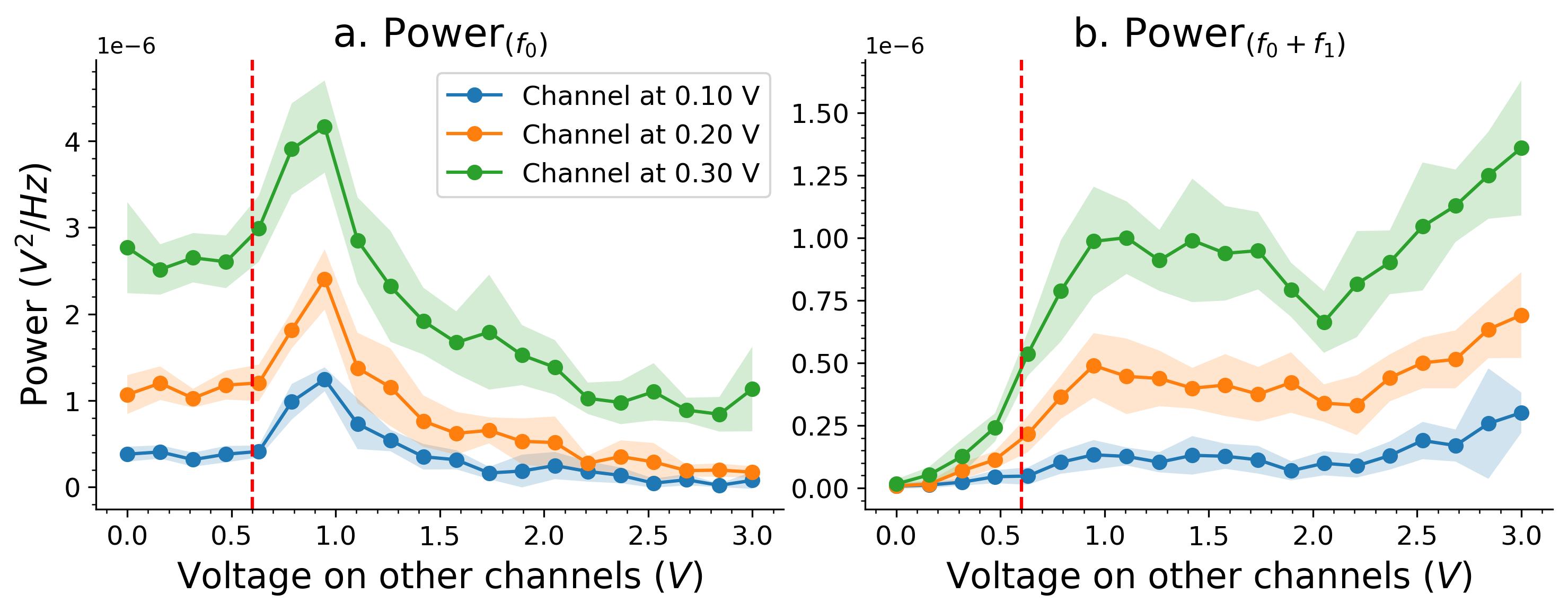}
    \caption{Analysis of the negligibility of crossed terms. a. represents the power at the frequency $f_0$ played in the optimized channel, b. represents the power at the frequency at the sum of frequencies between signals in different channels $f_0 + f_1$.}
    \label{fig:cross_terms}
\end{figure}

Figure~\ref{fig:cross_terms} shows the measured power spectral density at $f_0$ (a) and at the intermodulation frequency $(f_0 + f_1)$ (b) as a function of the voltage applied to the non-target channels. For voltages below approximately $0.6\,\mathrm{V}$, the power at $f_0$ remains approximately constant and no measurable component is observed at $(f_0 + f_1)$, consistent with the linear model. Above $\sim 0.6\,V$, a component at $(f_0 + f_1)$ becomes detectable and the amplitude at $f_0$ deviates from its low-voltage value. This behaviour indicates the presence of quadratic mixing between perturbations and defines the upper bound of the small-signal regime.

\section{Optimization Routines}

\subsection{Genetic Algorithm for Global Exploration}

The first stage of the optimization utilizes a genetic algorithm (GA) to perform a global exploration of the mask population. The algorithm begins by generating a population of random masks. Each individual mask is projected onto the DMD, and the resulting differential voltage signal is acquired to evaluate the fitness of the mask. The fitness function is designed to incentivize the isolation of specific sensing events. Through iterative cycles of tournament selection, crossover, and mutation, the system converges toward a robust initial guess that identifies coarse regions of the speckle field correlated with the target perturbations. In the paper, genetic optimization was performed over 20 generations, using a population size of 35 individuals, with each photodetector measurement timing at 100\,ms. 

\begin{algorithm}[H]
\caption{Genetic Algorithm for Global Mask Exploration}

\textbf{Input:}
\begin{itemize}
    \item $f$: target frequency
    \item $P$: population size
    \item $G$: number of generations
\end{itemize}

\textbf{Output:}
\begin{itemize}
    \item $\mathbf{W}_{GA}$: baseline optimal mask
\end{itemize}

\textbf{Procedure:}
\begin{enumerate}
    \item Generate an initial population of random masks
    $\{\mathbf{w}_1, \ldots, \mathbf{w}_P\}$.
    \item \textbf{for} $n = 1$ \textbf{to} $G$ \textbf{do}
    \begin{enumerate}
        \item \textbf{for} each mask $\mathbf{w}_i$ in the population \textbf{do}
        \begin{enumerate}
            \item Project $\mathbf{w}_i$ onto the DMD.
            \item Acquire the differential voltage $V_{\mathrm{diff}}$ from the photodetectors.
            \item Evaluate the fitness $\mathcal{F}$ using a signal-ratio metric.
        \end{enumerate}
        \item Select top-performing individuals as elites using tournament selection.
        \item Perform crossover to generate offspring from selected parents.
        \item Apply mutation to introduce genetic diversity.
        \item Update the population with elites and newly generated offspring.
    \end{enumerate}
    \item \textbf{return} the best mask $\mathbf{W}_{GA}$.
\end{enumerate}

\end{algorithm}

\subsection{Cross-Entropy Method for Local Fine-Tuning}

To perform a finer optimization, the final output of the genetic algorithm is used to initialize a Cross-Entropy Method (CEM). In this stage, each macropixel is treated as an independent probability distribution defined by a mean and a standard deviation. The algorithm samples a new population of masks from these distributions and evaluates their performance using the same fitness metric. By identifying the elite samples in each iteration, the probability parameters are updated to steer the distribution toward an optimal state. This fine-tuning stage reinforces the spatial weights associated with target frequencies, leading to an enhancement of the SNR of the desired signal. For the purpose of this article, we ran the algorithm for 20 generations, using a population size of 35 for each optimization, with the photodetector measurement spanning 100\,ms.

\begin{algorithm}[H]
\caption{Cross-Entropy Method for Local Fine-Tuning}

\textbf{Input:}
\begin{itemize}
    \item $\mathbf{W}_{GA}$: baseline mask from genetic algorithm
    \item $K$: number of iterations
    \item $P$: population size
\end{itemize}

\textbf{Output:}
\begin{itemize}
    \item $\mathbf{W}_{CEM}$: refined interrogation mask
\end{itemize}

\textbf{Procedure:}
\begin{enumerate}
    \item Initialize the probability distribution parameters $(\mu, \sigma)$ using $\mathbf{W}_{GA}$.
    \item \textbf{for} $n = 1$ \textbf{to} $K$ \textbf{do}
    \begin{enumerate}
        \item Sample a population of $P$ masks from the current distribution.
        \item \textbf{for} each sampled mask \textbf{do}
        \begin{enumerate}
            \item Project the mask onto the DMD.
            \item Acquire the differential voltage $V_{\mathrm{diff}}$.
            \item Evaluate the fitness $\mathcal{F}$.
        \end{enumerate}
        \item Identify the elite samples with the highest fitness scores.
        \item Update $(\mu, \sigma)$ toward the elite samples.
    \end{enumerate}
    \item \textbf{return} the final mean mask $\mathbf{W}_{CEM}$.
\end{enumerate}

\end{algorithm}

\section{Additional Results}

\subsection{Optimized Coefficients for Each Deformation Position}

Since it is not trivial to multiplex the training stage, we optimize one deformation channel at a time. In this section, we can see the loss curves for each deformation throughout the entire training routine, along with the final masks obtained.

\begin{figure}[H]
    \centering
    \includegraphics[width=\linewidth]{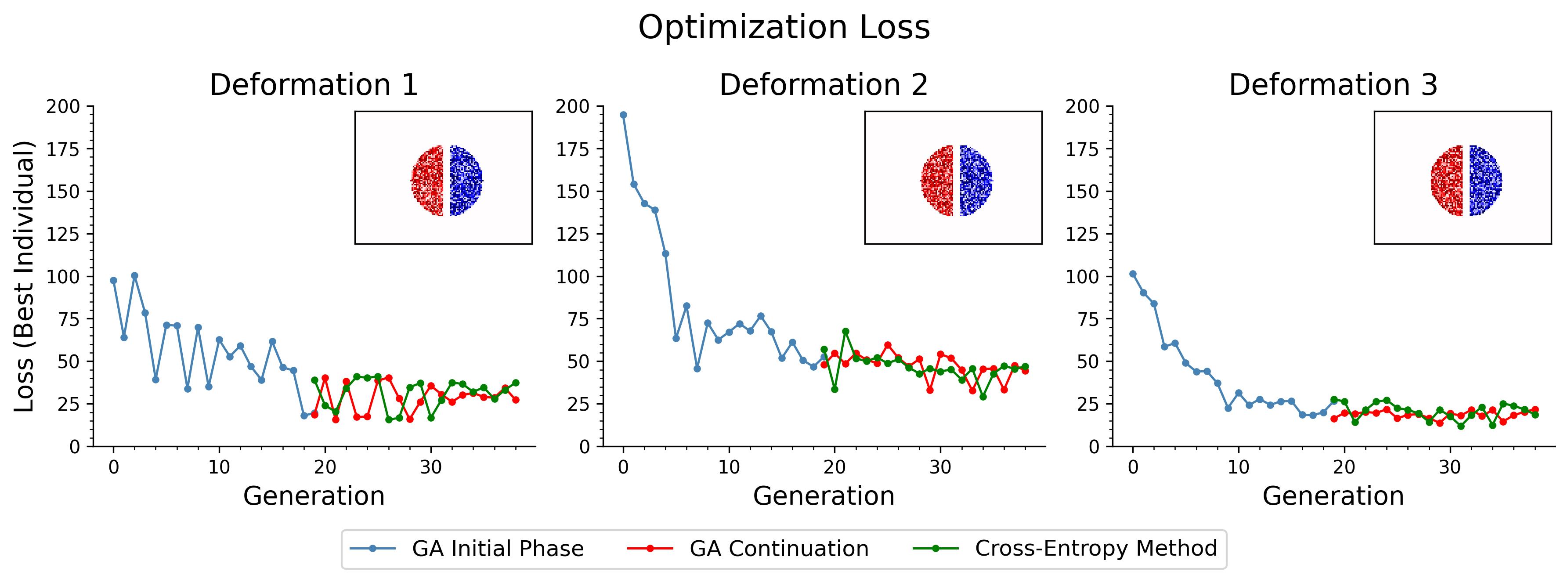}
    \caption{Loss curves and final optimized masks for each point of deformation. For illustrative purposes, we use red and blue to highlight the coefficients corresponding to positive and negative weights.}
    \label{fig:loss_curves}
\end{figure}

\subsection{Piezoelectric Membrane Characterization}

Throughout this work, piezoelectric membranes serve as the primary sources of physical perturbation. Since the modulation of the output speckle corresponds to the amplitude of the applied deformation, a characterization of the membrane displacement as a function of the input signal frequency was performed. This characterization utilized a Michelson interferometer, where the reference arm was controlled by a nanometer-resolution stage (MAX312D Stage, Thorlabs), and the probe arm consisted of a piezoelectric membrane attached to a mirror.

\begin{figure}[H]
    \centering
    \includegraphics[width=\linewidth]{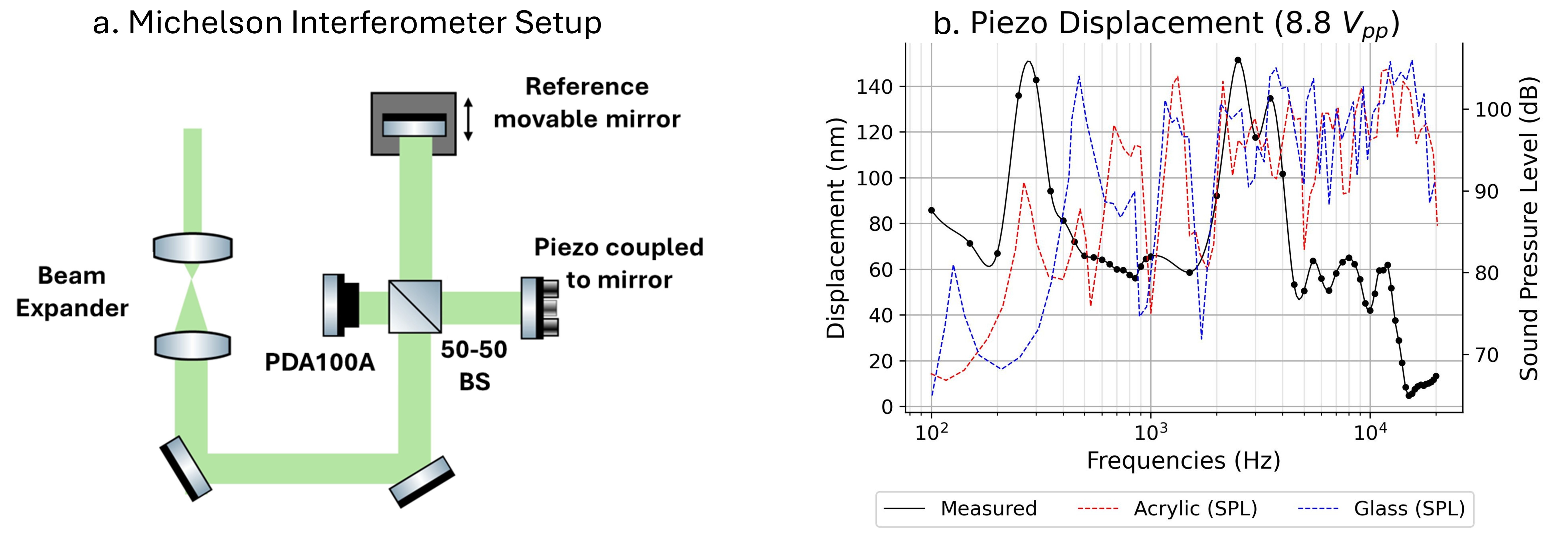}
    \caption{a. Schematic of the Michelson interferometer setup used in this manuscript. b. Displacement of the piezoelectric membrane according to the frequency of the applied signal. The data shown in black represent the experimental measurements performed with a Michelson interferometer. The red and blue lines show the sound pressure level measured for these membranes by the supplier, as described in the datasheet, where, in these cases, the piezo is glued to acrylic and glass plates.}
    \label{fig:piezo_characterization}
\end{figure}

Because the deformation of the membrane was smaller than the working wavelength of 532 nm, fringe counting was not viable. Consequently, the nanometer stage was used to identify the proper quadrature point of the interferometer, and, assuming a linear response regime, the deformation amplitude was extracted as seen in the displacement plot of Figure \ref{fig:piezo_characterization}. This figure also provides a comparison with the sound pressure levels specified in the component datasheet (PHUA3015-049B-00-000, TDK). As observed in the results, deformations above the 10 kHz regime exhibit a significant decrease in amplitude, making it difficult for such perturbations to couple into the optical fiber. For this reason, the modulation frequencies used in the experiments in this article were restricted to a range of up to 8 kHz.

\begin{figure}
    \centering
    \includegraphics[width=0.5\linewidth]{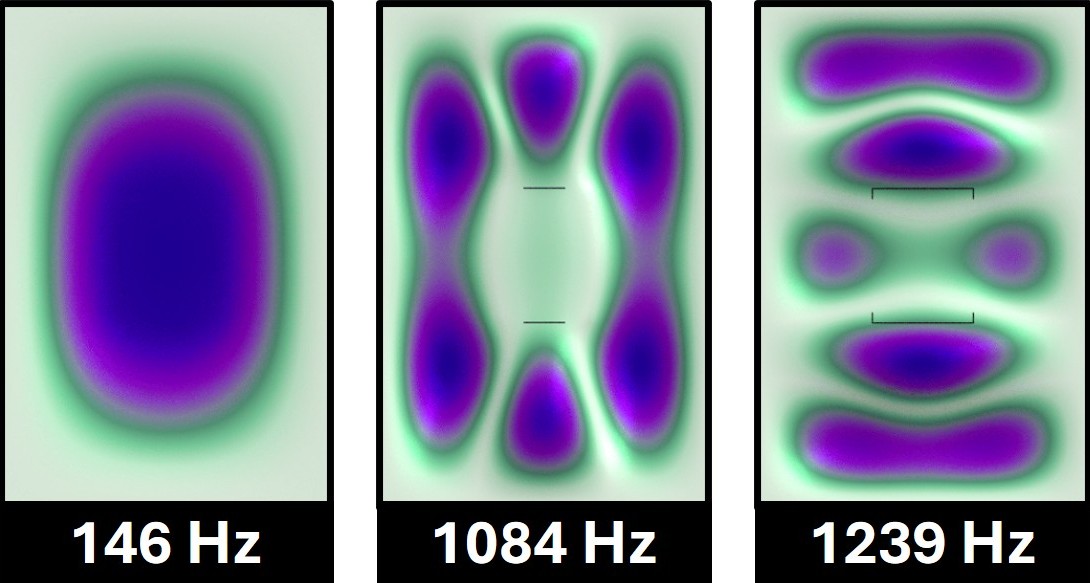}
    \caption{Image adapted from https://www.inventvm.com/piezodrive/, displaying the vibration modes across the surface of the piezo membranes for three distinct frequencies.}
    \label{fig:vibration_modes}
\end{figure}

Additionally, the different frequencies excite different vibration modes as can be seen in Figure \ref{fig:vibration_modes}. This causes additional difficulties in coupling all frequencies efficiently to the fiber.

Here, it is important to note that an unwanted effect arises from the frequency-dependent modal behavior of the actuator membranes, specifically that different regions of the membrane surface contact the fiber at different frequencies. This behavior implies that a spatial mask optimized for a single frequency may fail to maintain performance across the entire spectrum of interest. To ensure the system remains robust against these frequency-dependent contact regions, the optimization routine is performed using a frequency comb. This approach allows the evolutionary algorithm to identify spatial weights that are effective for the entire range of modulation frequencies simultaneously.

\subsection{Complex signature signals}

\begin{figure}
    \centering
    \includegraphics[width=\linewidth]{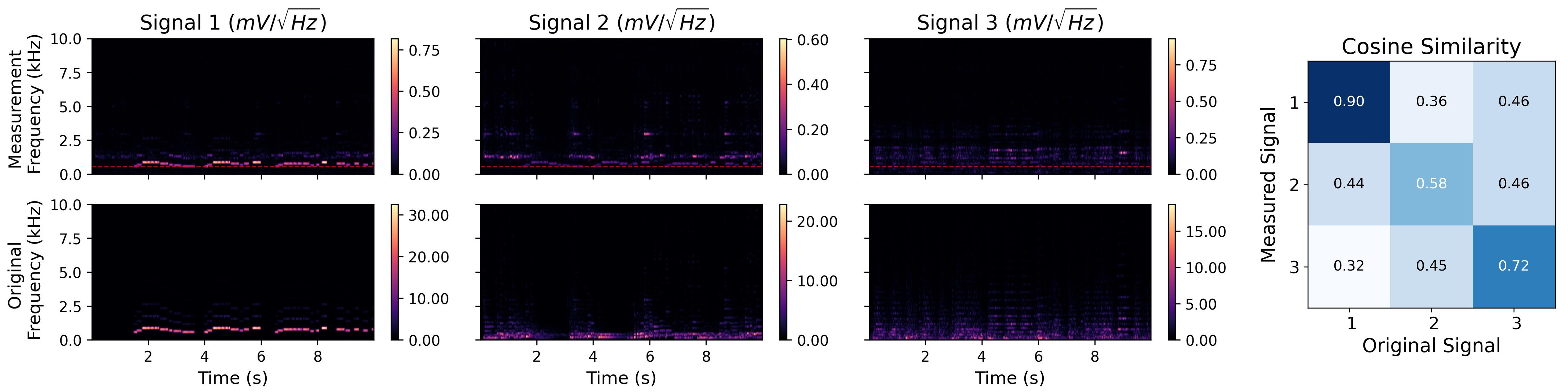}
    \caption{(Top) Spectrograms of the readout signals obtained when masks optimized for each deformation location are applied. (Bottom) Spectrograms of the three signals being applied simultaneously at three independent locations in the optical fiber. On the right, we compute the cosine similarity between each spectrogram for frequencies above 550\,Hz.}
    \label{fig:music}
\end{figure}

In addition to the results presented in the manuscript for the separation of three simultaneous signals (two frequency ramps and one constant frequency), we performed a similar demonstration where we applied three signals of higher spectral complexity. For this, we measured the readout differential signal across three different runs, each utilizing a mask optimized for a different deformation location. Figure \ref{fig:music} demonstrates these measurements in comparison to the signal playing at the target location. 

Here, similarities between the measurements and the targets are evident, with minor crosstalk present in some regions of the spectrogram. This is mainly observed for measurement location 2, where components from the signal at location 1 can be seen, although attenuated. To extract quantitative measures from these results, we also computed the cosine similarity between each spectrogram, targeting only frequencies above 550\,Hz. As previously mentioned, mechanical coupling difficulties prevented frequencies below this threshold from being efficiently translated into the optical fiber.

Audio recordings of the three measurements have been included as additional files along with the original clips for better comparison.

\backmatter
